\documentclass[12pt]{iopart}
\begin{document}

\title{From Einstein's Theorem to Bell's Theorem: \\ A History of Quantum Nonlocality} 

\author{H M Wiseman}

\address{Centre for Quantum Dynamics, School of Science, Griffith
University,  Brisbane,
Queensland 4111 Australia. }
\ead{H.Wiseman@griffith.edu.au}

\newcommand{\sch}{Schr\"odinger}
\newcommand{\hei}{Heisenberg}

\begin{abstract}
In this Einstein Year of Physics it seems appropriate to look at an important aspect of Einstein's work that is often down-played: his contribution to the debate on the interpretation of quantum mechanics. Contrary to physics `folklore', Bohr had no defence against Einstein's 1935 attack (the EPR paper) on the claimed completeness of orthodox quantum mechanics.  I suggest that Einstein's argument, as stated most clearly in 1946, could justly be called Einstein's reality--locality--completeness theorem, since it proves that one of these three must be false. Einstein's instinct was that  completeness of orthodox quantum mechanics was the falsehood, but he failed in his quest to find a more complete theory that respected reality and locality. Einstein's theorem, and possibly Einstein's failure, inspired John Bell in 1964 to prove his reality--locality theorem. This strengthened Einstein's theorem (but showed the futility of his quest) by demonstrating that either reality or locality is a falsehood. This revealed the full nonlocality of the quantum world for the first time. \end{abstract}



\maketitle

\section{Introduction}

It is a common view of Einstein, held even by sympathetic biographers, that his first 
two decades (1905 to 1925) of near-miraculous creativity
(including special and general relativity, stochastic methods, the photon 
and its interactions, and Bose-Einstein condensation) 
were followed by three decades which saw Einstein's `withdrawal from the 
contemporary problems of theoretical physics' \cite{Lan74} \footnote{The name of this book, {\em The Einstein Decade}, derives from its Part II, in 
which the author (a former research assistant of Einstein) gives synopses of all of  
Einstein's papers from 1905 to 1915. 
Part I (124 pages) is  essentially a scientific biography of Einstein, and is not
limited to these years. But Bose-Einstein condensation (1925) 
is the last scientific contribution discussed in any detail, and the entire debate with Bohr goes unmentioned.}. In this essay Einstein's later work is shown in a different light. 
At least in the period 1927--1935, he was at the forefront in the debate on the 
interpretation of quantum mechanics (QM). He abhored the nonlocality
in the Copenhagen interpretation, and thought it showed that the theory was incomplete.
That is, he thought that a theory with a realistic description of 
the microscopic world would eventually be found, and would eradicate the blight of nonlocality. 
This was the dream towards which he  failed to make progress for three decades.

Although Einstein failed, this does  
not invalidate his criticisms of the orthodox interpretation, 
most famously expressed in the 1935 Einstein-Podolsky-Rosen (EPR) paper
\cite{EinPodRos35}. Indeed,  I propose that the conclusion Einstein drew from the EPR 
 scenario, stated most clearly in 1946 \cite{Ein46} (p.~85), 
 could justly be called Einstein's theorem on reality, locality and completeness. 
 And there is a direct line from that theorem to the present field of quantum information science,
both theory and experiment \cite{PleVed98}.  (That is not to claim that there are not other lines of descent \cite{NieChu00,SpiMunBarKok05}.) 
It is primarily for this reason that the EPR paper, now one of Einstein's most cited papers, has received the overwhelming majority of its citations in the last 20 years \cite{Red05}. 
Indeed, only 20 years ago what are now called `entangled states' 
(a name coined by another quantum skeptic, \sch\ \cite{Sch35}) were almost universally   
known as `EPR-correlated states'. 

Einstein's theorem did not fully reveal the quantum nonlocality of entangled states which 
is central to quantum information science. 
But neither was this nonlocality discovered  by any of his opponents in the orthodox camp. 
Rather,  it was discovered by one of his followers, 
 John Bell, by means of his 1964 theorem on reality and locality \cite{Bel64}. 
 Bell's theorem actually implies Einstein's theorem, 
 but it also implies that Einstein's 
 quest to find a local realistic theory was doomed from the start. Perhaps  Einstein's
 grand failure in this quest may have suggested to Bell that 
 such a theory might not exist. In any case, although Bell's theorem destroyed Einstein's dream  
it laid the foundations for our modern understanding of the quantum world. 

This essay traces the development of ideas about locality and reality in QM from 
1927 to 1964 and beyond. Einstein and Bell loom largest here, but more by desert than by design. 
Bell would have featured even more prominently if I had presented a proof of his theorem, but that 
would have made this essay too long. For those who are not familiar with Bell's theorem, 
Bell's original paper \cite{Bel64} is quite readable, and in addition 
numerous authors (including me) have written popular accounts --- see Refs.~\cite{Har98,JacWis05} and works cited therein. Even for those who are familiar with Bell's theorem, I hope this essay will provide new light on its historical origins and significance. 

This essay is dedicated to my past students in the Griffith subject  
{\em Life, The Universe, and Everything}, for expressing thoughts that,
from time to time, have forced me to re-evaluate  my understanding of much of this material.

\section{1927: Bohr, Heisenberg, de Broglie, and Einstein} \label{sec:BHDE}

In the second quantum revolution of the mid 1920s, almost all of Bohr's
old quantum theory, with electrons jumping between stationary atomic orbits,  
was abandoned. Perhaps for this reason, Bohr
was strongly committed to indeterminism, as he argued that
his idea of quantum jumps was still an essential part of 
QM \cite{Bel99}. Heisenberg, on the other hand, wanted completeness,
because he wanted to be seen as the principal author of an entirely
new world view \cite{Bel99}. Other major players, including
 Schr\"odinger, Einstein, and de Broglie, were convinced
 neither of indeterminism nor completeness.

 To strengthen their  positions, Heisenberg and Bohr constructed a united public
stance which became known as the Copenhagen
interpretation  \cite{Bel99}, combining \hei's uncertainty relations \cite{Hei27} with Bohr's principle
 of complementarity \cite{Boh28}. Heisenberg argued that it was impossible to know 
both the position and momentum of a particle because the apparatus disturbed it, 
while Bohr interpreted this by saying that they were complementary properties, 
 which could not be measured by the same apparatus. 
Bohr and \hei\  concluded that the properties of quantum systems did not exist independently of the apparatus. Rather, the values obtained in an experiment were created by the apparatus, with 
the  probabilities for the possible results determined by the wavefunction $\psi$. 
 Thus the wavefunction both affected the apparatus and was affected by it 
 (the quantum jump upon measurement). 
This version of QM was introduced 
 by Bohr at the September 1927 meeting in Como. 
 
From the start Einstein was troubled by the fact that these quantum jumps 
violated local causality (often abbreviated to locality). This is the 
requirement from special relativity that an 
event cannot be influenced by other events that are space-like separated. At the October 1927 Solvay conference Einstein illustrated this nonlocality by considering the collapse of the wavefunction of a single particle \cite{Wic95}. Since this wavefunction 
could become arbitrarily spread-out over time, the detection of the particle by an observer at
one location would, in the Copenhagen interpretation, instantaneously change the wavefunction 
over all space in order to prevent other (arbitrarily distant) observers from also observing that particle at their location. 

The point of the thought-experiment was of course that the mysterious action at a distance would vanish if one were to defy Heisenberg's and Bohr's insistence that $\psi$ was a complete description of the particle's state, and 
instead allow that the particle had a real position at all times. Then detecting (or not) the particle at one location would simply reveal that the particle was (or was not) present  there, and hence absent (or present) elsewhere. A pre-existing  position was a feature of the `pilot wave' theory that de Broglie 
proposed in 1927 \cite{deB56}, and at this Solvay conference Einstein said: \cite{Jam74}
\begin{quote}
It seems to me that this difficulty [action at a distance]
cannot be overcome unless the description of the
process in terms of the Schr\" odinger wave is supplemented
by some detailed specification of the localization of the
particle during its propagation. I think M. de Broglie
 is right in searching in this direction. If one
works only with Schr\"odinger waves, the [Copenhagen] interpretation of $|\psi|^2$, 
I think, contradicts the postulates of relativity.Ó
\end{quote}

Why was Einstein not {\em more} enthusiastic about de Broglie's theory? 
The answer is probably that he was aware that de Broglie's theory had its own problem of nonlocality. 
This was not as gross as the nonlocality in Copenhagen QM, as it was present only for a system with more than one particle. In such systems the wavefunction is not a function in real space, but rather in 
 multi-dimensional configuration space. In de Broglie's theory, the trajectory of one particle 
could be influenced by the position of another particle, even if the two were distant and non-interacting \cite{deB56}. 

Einstein would have been familiar with these facts because he had been pursuing a similar idea himself, but abandoned it when its nonlocality became evident \cite{Cus94,Bel97}. 
Thus, although in 1927 Einstein was still promoting the idea of particle trajectories, his own research had convinced him that any approach based on a wavefunction in multi-dimensional configuration space was bound to be nonlocal. 
Indeed, the quote above continues with reference to  `objections of principle against this multi-dimensional representation [of reality]' \cite{Bel97}. 

The Copenhagen-adherents  at the 1927 Solvay conference were, not surprisingly, 
dismissive of de Broglie's approach. However, the criticisms leveled at it, especially by Pauli, 
do not hold up to scrutiny \cite{Val01}. Nevertheless they, and Einstein's lack of real enthusiasm, were enough to make de Broglie abandon his research program. Interestingly, no criticism based
on the theory's nonlocality is recorded. Perhaps it was noticed but went unremarked upon because the Copenhagen interpretation was nonlocal even in the single-particle case. Indeed, Heisenberg admitted this shortly afterwards \cite{Hei30}, using an example like Einstein's, but with the wavefunction split into parts (transmitted and reflected) by a beam splitter. He said:
\begin{quote}
The [measurement] at the
position of the reflected packet thus exerts a
kind of action (reduction of the wave packet)
at the distant point occupied by the transmitted
packet, and one sees that this action is
propagated with a velocity greater than that
of light. However, it is also obvious that this kind of action can
never be utilized for the transmission of signals so that
it is not in conflict with the postulates of the theory of
relativity.
\end{quote}

As indicated by the earlier quote, Einstein had the opposite view regarding the compatibility of this nonlocality with the postulates of the theory of relativity. Einstein's view here is surely the correct one, since he was the author of these postulates!  The impossibility of superluminal signalling is merely a {\em consequence} of the postulates of the theory of relativity. The postulates themselves are much stronger. As Einstein wrote in 1905  \cite{Ein05}, 
`the velocity of light in our theory plays the part,
physically, of an infinitely great velocity'. It could scarcely be clearer that 
the foundations of the  theory of relativity are in direct conflict with \hei's `action [that] is
propagated with a velocity greater than that of light'. 

So skeptical did Einstein become of QM at this point that he thought he could 
prove that it was inconsistent, starting at that eventful 1927 Solvay meeting.   
He thought he had found an 
experiment that would enable one to determine simultaneously the
position and momentum of a particle. But Bohr rose to the challenge, showing that the
uncertainty principle applied to the apparatus would be the undoing
of all of Einstein's thought-experiments \cite{BohrEinst}.
 In this debate Bohr was essentially correct;  his semiclassical arguments  
 would have been completely rigorous if the system and apparatus were
 to have been in a state with a positive Wigner function  \cite{Wig32}, which can be treated as a joint 
  probability distribution for the positions and momenta.  
  The irony of this is that such semiclassical arguments 
are based on a picture of underlying reality whose existence Bohr denied! But this was
probably due more to Bohr's lack of mathematical ability than inconsistency 
in his world view \cite{Bel99}.

\section{1935: Einstein, Podolsky, and Rosen} \label{sec:EPR}
  
  After 1931 Einstein appeared to accept
  that he had been wrong, and that QM was consistent \cite{Pai82}. 
  Indeed, in later life he went so far as to say that
`The entire mathematical formalism [of QM] will probably have
to be contained, in the form of logical inferences, in every useful future theory.'
\cite{Ein49} (p.~667). However this did not mean that he accepted Bohr's {\em interpretation} of QM,
and   his next attack was to become the most famous. 

 The 1935 Einstein-Podolsky-Rosen (EPR) argument \cite{EinPodRos35}
  was meant to prove that Copenhagen QM was incomplete. (Fine \cite{Fin96} and Norsen \cite{Nor05} emphasize that Einstein's views should not be equated with the argument in the EPR paper. I discuss this 
  in the Appendix.) 
EPR considered a pure state describing two particles which are well-separated but   
correlated. This is the sort of state to which \sch\  
was shortly to give the name `entangled' \cite{Sch35}.
 The particular entangled state EPR considered had the properties that 
if one were to measure the position of particle $S_1$, one could infer with arbitrary accuracy 
the position of particle $S_2$, and the same for the momenta. (These positions and momenta could be taken to be in a direction orthogonal to that separating the two particles, so that any uncertainty in these positions does not affect the fact that the particles are well-separated.)

 EPR began by defining `element of physical reality'   (which I will not abbreviate as EPR)
in a way that cannot credibly be criticized: 
\begin{quote}
If, without in any way disturbing a system, we can predict with certainty \ldots\ the value of a physical quantity, then there exists an element of physical reality corresponding to this physical quantity.
\end{quote}
 In the EPR state a distant observer can find out the position of $S_2$ by a measurement on $S_1$. {\em By locality}, such a measurement cannot in any way disturb $S_2$. Thus, by EPR's definition, it follows that the position of $S_2$ is an element of physical reality. The same argument holds for the momentum of $S_2$.   But Copenhagen QM says that the position and momentum of $S_2$ cannot both be determinate, and in particular for the EPR state ascribes to {\em each} of them an arbitrarily large uncertainty. Therefore, EPR concluded, Copenhagen QM is incomplete.  

The alternative, as EPR said, would be to make the `reality [of the second system] \ldots\ depend upon on process of measurement carried out on the first system.' They scorn (but do not logically exclude) this nonlocality, saying that `No reasonable definition of reality could be expected to permit this.'  The EPR scenario itself suggests that the Copenhagen interpretation is incomplete, rather than that the world really is nonlocal. As for the single particle, the EPR paradox vanishes if one supplements QM by  `hidden variables' (HVs). In this case, the HVs should  correspond to the particles' positions and momenta. This works because the EPR state has a positive Wigner function, so the Wigner function variables $(q_1,p_1,q_2,p_2)$ can act as HVs.
 
 One might wonder why EPR chose to introduce their two-particle argument in the first place, since the same conclusions also follow from Einstein's much simpler single-particle argument  \cite{Nor05}. An important improvement offered by the EPR scenario is the following. Consider \hei's two-wavepacket version of the single particle scenario. It is crucial that the state of the single particle be a coherent superposition of the   two wavepackets. If the state of the particle were a mixture of being  in the reflected wave or the transmitted wave, then there may be no mystery as to why the particle turns up only at one place --- even a believer of Copenhagen QM could claim that the particle was actually in one wavepacket or the other. 
 (Note that they would have to believe that the mixture is a {\em proper} mixture, in the sense of d'Espagnat \cite{dEs89}.)  But to verify that the state of the particle was a coherent superposition it would be necessary to bring the wavepackets together again to demonstrate interference. By contrast, in the EPR scenario the correlations between the results of {\em local} measurements of position and momentum \cite{Rei89} are all that is required to realize the paradox experimentally. Although Einstein may have regarded the EPR paradox purely as a thought-experiment, \sch\ certainly thought that it was important to test such predictions experimentally, as I will discuss in Sec.~\ref{sec:SVB}. 
 
\section{1935: Bohr's Reply} \label{sec:Bohr} 
 
For a long time it was thought (and possibly is still thought) by most physicists that Bohr's response \cite{Boh35} to EPR won the debate for the Copenhagen school.
Propagandists from this school portrayed Bohr's supposed victory in glowing terms \cite{Ros64}: `Einstein's problem was reshaped and its solution reformulated with such precision that the weakness in the critics' reasoning became evident, and their whole argument, for all its false brilliance, fell to pieces'. The truth is that EPR had presented a logical argument,  whereas Bohr's reply was a quagmire from which even his supporters had difficulty extracting any clear meaning  \cite{Bel99,Wic95}. (It is telling that when his reply was reprinted in the 1983 compendium by Wheeler and Zurek \cite{WheZur83}, no-one noticed prior to publication that the pages were printed out of order.)
 
  In the greater part of his reply, Bohr simply ignored the EPR set up, and
reiterated his old-style defence of the consistency of QM,  in which Einstein's attacks had been thwarted by invoking the disturbance of  the system by the apparatus. But these arguments  were now irrelevant because the apparatus for $S_1$ could not physically disturb $S_2$. When he did finally address the EPR set-up, he again saw his task as being to prove that  QM was consistent,  
this time because complementarity ensured that 
the position and momentum measurements of $S_1$ could not be done
simultaneously. This was also irrelevant because EPR were not questioning the consistency of QM.  
Indeed, the very title of the EPR paper, which Bohr's reply also bore, 
 was the question of completeness, not of consistency.  
 
 When reviewing the Einstein--Bohr debates in 1949 \cite{BohrEinst}, Bohr
 concluded his summary of his reply to EPR by quoting his defence based upon complementarity. 
Astonishingly, he  immediately
  followed this by an apology for his own `inefficiency of expression which must have made it very difficult to appreciate the trend of the argumentation \ldots'!
But rather than taking the opportunity to explain himself more lucidly, 
he instead directed the reader to concentrate on the earlier debates with Einstein
regarding the consistency of QM. 
  
Discounting the irrelevancies, what is left of Bohr's reply? 
Essentially just a restatement of the paradox \cite{Boh35}:  
  \begin{quote}
In  [EPR's] arrangement, it is therefore
clear that a subsequent single measurement
either of the position or of the momentum of
one of the particles will automatically determine
the position or momentum, respectively, of the
other particle with any desired accuracy \ldots\ [even though] there 
is no question of a mechanical disturbance of the system.
\end{quote}
This automatic determination without mechanism was precisely the 
`telepathy' \cite{Ein46} that Einstein objected to in the Copenhagen interpretation. 
While Bohr did not directly admit that his interpretation was nonlocal, he did sometimes 
use words like `wholeness' \cite{Boh63} that arguably \cite{dEs84,Cus94b} 
amount to much the same thing. 

\section{\sch, von Neumann, and Bohm} \label{sec:SVB}

Schr\"odinger \cite{Sch35,Sch35b,Sch36} also made major contributions to the debate following the EPR paper. He pointed out again the nonlocality of the Copenhagen interpretation when dealing with 
`entangled' states as he called them \cite{Sch35}. He also coined the term
 `steering' \cite{Sch35b} or `driving' \cite{Sch36} for the EPR phenomenon 
 where measurements on one systems directly influence the state of a  distant system.  
While Einstein  thought that Copenhagen QM was incomplete, \sch\ actually doubted that it was 
correct in its description of spatially distant systems \cite{Sch36}. That is, \sch\ doubted that 
the EPR correlations could be seen experimentally.  
Finally, he highlighted the quantum measurement problem: the fact  that there was nothing 
in the theory that would prevent microscopic unreality infecting the macroscopic world. 
This was, of course, the infamous \sch's cat paradox \cite{Sch35}, that showed how 
the cut required by Copenhagen QM between the quantum and 
classical world was ill-defined and unsatisfactory. 

One might have expected the arguments by EPR and \sch\ to have 
 led to renewed interest in theories that sought to complete QM, since such theories  
 offered the prospect of solving both the nonlocality problem and the quantum measurement problem. 
 However, by this time popular support for the Copenhagen interpretation over any `hidden variable' theory had been bolstered by von Neumann's supposed proof of the 
impossibility of HVs published in 1932 \cite{Von32}. The effectiveness of the Copenhagen school's 
use of von Neumann's authority to silence its critics is discussed in Ref.~\cite{Fey95}. 

As is now well-known, von Neumann's proof, while technically correct, 
made unwarranted assumptions about the nature of HVs which completely undermine
its claim to be an impossibility theorem.  This flaw  was 
pointed out in 1935 by Grete Hermann in an obscure philosophy journal \cite{Wic95}, 
and around 1938 Einstein independently showed the flaw to his
assistant Bargmann (according to the latter)  \cite{Wic95}. 

In 1952 David Bohm published his own HV interpretation \cite{Boh52}, having
rediscovered de Broglie's idea. In de Broglie's formulation, the HVs were the position of particles, but Bohm and co-workers eventually broadened this to include the values of gauge fields at all points in space \cite{BohHil93}. 
Bohm presented a comprehensive account of his theory, showing its consistency and  its 
superiority to the Copenhagen interpretation in solving the quantum measurement problem.

Again, one might have expected Bohm's HV theory to have exposed the flaw in 
von Neumann's theorem (although Bohm himself failed to identify the flaw). Instead, most
physicists assumed that Bohm's theory must be incorrect \cite{Whi96}. Einstein of course knew better, 
but he still thought the solution `too cheap'   \cite{Whi96}. Bohm's theory 
used the wavefunction in configuration space as a guiding wave, which, 
Einstein rightly saw, was bound to lead to nonlocal behaviour.
Einstein was evidently still hoping that a more radical theory could restore locality.

Ironically, just prior to developing his HV intrepretation, Bohm had defended the Copenhagen interpretation in his well-regarded 1951 text book \cite{Boh51}. As it would turn out, 
probably the most influential aspect of this book was that, when discussing the EPR paradox (Chapter XXII), Bohm introduced a simplified   scenario involving two spin-half particles with correlated spins, rather than two particles with correlated positions and momenta as used by EPR. As well as being simpler to analyze, the EPR-Bohm scenario has the advantage of being experimentally accessible. Indeed, as Bohm and Aharonov pointed out in 1957 \cite{BohAha57}, the experiment had already been done in 1950  (for very different motivations), verifying the predictions of QM. 

\section{1946: Einstein's Theorem} \label{sec:ET}

At age 67 (in 1946 or early 1947), Einstein wrote a short scientific autobiography \cite{Ein46} in which he set out his views on the foundations of physics in depth. It is here that one finds the clearest expression of what can be deduced from the EPR phenomenon, which \sch\ called steering. Einstein referred to `the most successful physical theory of our period, viz. the statistical quantum theory'. By this Einstein meant the minimal quantum theory in which $\psi$ has no interpretation except as a means to answer questions of the form: 
`What is the probability of finding a definite physical magnitude $q$ (or $p$) in a definitely given interval, if I measure it at time $t$?' \cite{Ein46} (p.~83).  [Here `$q$ (or $p$)' should be understood to imply arbitrary observables.] Having succinctly explained the EPR paradox, he said `physicists [who] accept this consideration \ldots\ have to give up [the] position that the $\psi$-function constitutes a complete description of the real factual situation'. \cite{Ein46} (p.~85). None of this appears much  different from Einstein's 1935 arguments. But here for the first time he also stated:  \cite{Ein46} (p.~85)  
\begin{quote}
  One can escape from this conclusion [that statistical quantum theory is incomplete] only by either assuming that the measurement of $S_1$ (telepathically) changes the real situation of $S_2$ or by denying independent real situations as such to things which are spatially separated from each other. Both alternatives appear to me equally unacceptable.
  \end{quote}
  
 Omitting the opinion (clearly stated as such) about what was acceptable in a physical theory,
 the logical deduction to which Einstein came in 1946 was that  one of the following is false:
\begin{enumerate}
\item the completeness of statistical QM
\item locality (that is, the postulates of relativity)
\item the independent reality of distant things.
\end{enumerate}     
This result could, in my opinion, justly be called Einstein's reality--locality--completeness theorem, 
so that it can stand alongside Bell's reality--locality theorem \cite{Bel64}. (Einstein would not necessarily have endorsed this term, since he disapproved of too much formalism \cite{Fin96}.) Recently, Norsen \cite{Nor05,Nor04} has set about presenting Einstein's arguments more formally, but he does not consider the third possible falsehood that Einstein listed in 1946, the existence of `real situations [for] things which are spatially separated'. 

In allowing for the possibility that distant things were not real, Einstein's 1946 theorem goes beyond his 1935 arguments addressing the Copenhagen interpretation (although see Ref.~\cite{How85} for a contrary view). Copenhagen QM, at least as Bohr conceived it \cite{Bel99}, presumed the existence of the classical world. In particular, it presumed that observer $A$ for system $S_1$ and observer $B$ for system $S_2$ were real, no matter how far apart they were. This presumption is why it can be proven to be a nonlocal theory. Statistical QM, as Einstein defined it, makes no such presumption. I suggest that Einstein was quite deliberately not assuming the existence of more than one observer when he said that statistical QM was about the value an observable may take `if {\em I} measure it' [my emphasis]. Statistical QM as a theory with a single observer $A$  allows $\psi$ to be interpreted as $A$'s knowledge of the quantum world. The fact that this changes instantaneously now poses no mystery, because it is merely a change in $A$'s knowledge. But precisely because it is only a change in $A$'s {\em knowledge}, this change in $\psi$ cannot {\em affect} the result $B$ obtains, even when (as with the EPR state) the results that $A$ and $B$ obtain may be perfectly correlated.  Statistical QM as a theory of $A$'s knowledge evades this problem, however, simply by denying that $B$ has an independent existence. All that exists ({\em if the theory is complete}) is $A$'s knowledge of $B$.

Perhaps regretting his 1930 admission of the nonlocality of the Copenhagen interpretation 
(see Sec.~\ref{sec:BHDE}), \hei\ by 1958 had been seduced by the idea that `the discontinuous change in our knowledge in the instant of registration \ldots\ has its image in the discontinuous change of the probability function [$\psi$]'. \cite{Hei58}.  But he failed to take the necessary logical step of accepting that $\psi$ can represent only one individual's knowledge. In contrast to Einstein's deductions, his famous statement \cite{Hei58b} that `In the Copenhagen interpretation of quantum mechanics, the objective reality has evaporated, and quantum mechanics does not represent particles, but rather, our knowledge \ldots\ of particles'. is beset with inconsistencies. If objective reality has evaporated, how can there exist a community of individuals who have knowledge of particles? How can QM describe {\em our} knowledge, since my knowledge does not change when you make an observation? Any  attempt to answer these questions consistently would lead either to Bohr's Copenhagen interpretation with its classical realm, or to extreme subjectivism, denying that there are  matters of fact about distant observers. \footnote{One can perhaps avoid subjectivism by being totally impartial, and denying that there is a matter of fact even about one's own experience. This was the route taken already in 1957 by Everett \cite{Eve57}: the relative-state or many-worlds interpretation.}
Thus the three options which Einstein's theorem gives us, corresponding to the three possible falsehoods above, are 
\begin{enumerate}
\item a hidden variables theory (which Einstein thought could be local)
\item Bohr's Copenhagen interpretation (which is nonlocal)
\item extreme subjectivism (which is local insofar as it denies the reality of distant events).
\end{enumerate} 

\section{1964: Bell's Theorem} \label{sec:Bell}

Unlike most physicists of the post-war generation, John Bell took Einstein's theorem seriously, 
quoting Refs.~\cite{Ein46,Ein49} to support the suggestion `that the hidden variable problem has some interest' in the introduction of his 1966 review article on this topic \cite{Bel66}. Bell had read Bohm's HV theory while still a student and this led him to rediscover the flaw in von Neumann's impossibility proof (see Sec.~\ref{sec:SVB}) \cite{Wic95}. But he did not write it up until 1964 \cite{Bel66}, and then publication was accidentally delayed for two years. In this review paper Bell also went beyond this discovery, posing the question as to whether all realistic theories must be, like Bohm's, nonlocal.
Then in the same year, having done the first 
complete analysis of correlations in the EPR-Bohm scenario (see Sec.~\ref{sec:SVB}), 
 he was able to answer his own questions `yes' \cite{Bel64}.
 
Einstein's theorem \cite{Ein46} showed that, if statistical QM is complete,
either the world is nonlocal or reality is restricted to my location. Bell proved that even if statistical QM is not complete, this does not allow local realism  --- HV theories (which by their nature are realistic) must also be nonlocal. Thus, adding Bell's result to Einstein's result \cite{Nor04}, the list of possible falsehoods is reduced to two:
\begin{enumerate}
\item locality (that is, the postulates of relativity)
\item the independent reality of distant events.
\end{enumerate}     

It has been argued \cite{Sta71,Ebe77} that this conclusion can also be drawn by 
reinterpreting Bell's proof, without appealing to Einstein's theorem. The reason is that the 
 `local hidden variables'  $\lambda$ that Bell proved to be inadequate  
 are simply the most general way to represent a local cause  for spatially separated measurement events (assumed real). Considering such variables does {\em not} presume that statistical QM is incomplete. For example, if $\psi$ factorizes as $\psi_A\otimes \psi_B$, then statistical QM {\em does} provide a local cause for separated measurement events. In this case $\psi_A$ and $\psi_B$, plus some random numbers associated with the act of measurement, can act as Bell's $\lambda$. If statistical QM were also to provide a local cause even for the case when $\psi$ is entangled, then Bell could not have proved his result. That is, Bell's proof stands by itself as a reality--locality theorem, and actually {\em implies} Einstein's theorem. 
 
Irrespective of this argument, it seems justified to attach Bell's name to the above reality--locality theorem. Bell was certainly aware of the significance of Einstein's theorem. 
In the second sentence of his paper \cite{Bel64}, he quoted Einstein \cite{Ein46} (p.~85) in a footnote, to make the point that statistical QM (that is, QM {\em without} HVs) violates locality or reality. Since Bell's result ruled out local HV theories, he concluded  \cite{Bel81}
that in an EPR-Bohm experiment, `we {\em cannot} dismiss intervention [by an experimenter] on one side as a causal influence on the other'.  The only alternative to accepting such nonlocality is for one experimenter to deny the independent reality of the other's experience. 
\footnote{To be scrupulous, there are perhaps four other ways that the correlations in such an experiment could be explained away.
(1) One could simply `refuse to consider the correlations mysterious' \cite{Fra82}.
(2) One could deny that the experimenters have free will to choose the settings of their measurement devices at random, as required for a statistically valid Bell-experiment \cite{Sta71}.  
(3) One could entertain the idea of backward-in-time causation \cite{Pri96}. 
(4) One could conclude that ordinary (Boolean) logic is not valid in our Universe \cite{Bub74}. I do not consider these escape routes because they seem to undercut the core assumptions necessary to undertake scientific experiments. Bell
expressed similar sentiments: With regard to option (1) he said `Outside [the] peculiar context  [of quantum philosophy], such an attitude would be dismissed as unscientific. The scientific attitude is that correlations cry out for explanation'. \cite{Bel81}. With regard to option (2) he thought it was not worth considering unless it could be shown to have some theoretical justification:  `When a theory \ldots\ in which such conspiracies inevitably occur \ldots\ is announced, I will not refuse to listen \ldots' \cite{Bel77}. In Bell's opinion, option (3) was the same as option (2): `I have not myself been able to make sense of backward causation. When I try to think of it I lapse quickly into fatalism', as quoted in Ref.~\cite{Pri96}. Finally, of option (4), Bell said that `When one remembers the role of the apparatus, ordinary logic is just fine'. \cite{Bel90}, and thought that a `full appreciation of this [role] would have aborted \ldots\ most of ``quantum logic'''. \cite{Bel87} (p.~vii). } 
However unpalatable this second option is, 
it also cannot be dismissed; as Bell said \cite{Bel81a},  `Solipsism cannot be refuted'. 

There are many ironies associated with Bell's 1964 paper. The first irony
is that, a generation after the Einstein--Bohr debates, 
physicists had forgotten (or, like \hei, convinced themselves that they could deny) 
 the nonlocality of the Copenhagen interpretation. 
Thus the community seemed \cite{Wig76} (and perhaps still seems \cite{Mer93}) 
to miss the point and to think that Bell's theorem showed that HV interpretations   
must be nonlocal {\em in contrast} to the Copenhagen interpretation \cite{Nor04}. 
Indeed, the journal editor who 
accepted Bell's paper, the renowned condensed-matter theorist Philip Anderson, did so partly because he thought that it refuted Bohm's theory \cite{Wic95}. This misunderstanding also contains the second irony; in showing that HV interpretations must be nonlocal, Bell's theorem in fact {\em nullified} any criticisms of Bohm's theory based upon its nonlocality.  Bell himself said that  `It is a merit of the
de Broglie--Bohm version to bring [the nonlocality of QM] out so explicitly that it cannot be ignored' \cite{Bel80}.  And this quote reveals the final irony; Bell, the destroyer of Einstein's dream of local realism, was, like Einstein, a vigorous opponent \cite{Bel90} of the Copenhagen interpretation.

Both Einstein's theorem and Bell's theorem as stated here 
assume the predictions of QM to be accurate. Any verification of the predictions that underpin Bell's theorem would also suffice to verify the predictions that underpin Einstein's theorem, but most experimentalists remained uninterested in testing QM in this way. However, there were a few persistent individuals \cite{Wic95}, and now Bell-experiments have been accepted as an interesting and important endeavour. The landmark Bell-experiment verifying the QM predictions is generally identified as that performed in 1981 by Aspect, Grangier, and Roger~\cite{AspGraRog81}. For a review of experiments up to 1987, see Ref.~\cite{Red87}, and for a more recent review of theory and experiment, see Ref.~\cite{Har98}.
 
If one assumes the world to be real, then Bell-experiments have proven that it is nonlocal. 
The nonlocality demonstrated in these experiments  does not enable superluminal signalling 
(and does not allow a preferred reference frame to be identified). 
 It has therefore been called {\em uncontrollable nonlocality} \cite{Shi84}, to contrast with (hypothetical) {\em controllable nonlocality} which would enable superluminal signalling \footnote{Shimony \cite{Shi93} has also used the phrase  `passion at a distance',  as opposed to `action at a distance', as a colourful way to explain uncontrollable, as opposed to controllable,  nonlocality. Other terms have also been used \cite{Jar84}. Finally it is interesting to note that Einstein's pejorative term `telepathy' \cite{Ein46} has been recently resurrected (modified to `quantum pseudo-telepathy') \cite{Bra04} for the cases where the usefulness of uncontrollable nonlocality is most evident.}. 
But this uncontrollable nonlocality 
is not purely notional. It can be used to perform tasks \cite{Bra04,JacWis05} which would be  impossible in a world conforming to the postulates of relativity \cite{Mau94}.  Thus, uncontrollable nonlocality reduces the status of relativity theory from fundamental to phenomenological \cite{Bel87} (p.~172).
  
On the other hand, if one assumes that relativity is fundamental, then Bell-experiments have proven that there is no real world. (To be more precise, under this assumption  
Bell-experiments have proven that  the real world exists only within one's own past light-cone \cite{Ken05}). It cannot be stressed enough that this does
{\em not} mean merely that microscopic particles or fields cannot  
have determinite properties. It means macroscopic objects, other conscious observers even, 
are not real at the present time. 
That is why Stapp \cite{Sta77} has called Bell's theorem `the
most profound discovery of science.'

To conclude,  the results of the Bell-experiments leave only two
 possibilities:
 \begin{enumerate}
\item the world is nonlocal --- events happen which violate the principles of relativity
\item objective reality does not exist --- there is no matter of fact about distant events. 
\end{enumerate} 
Although consistent, the second option does seem very close to solipsism, 
`the view or theory that only the self really exists or can be known' \cite{OED}. 
Even if solipsism cannot be refuted, it can certainly be attacked on ethical grounds \cite{Bel81a}. 
 As Karl Popper wrote pointedly  \cite{Pop85},  `any argument against realism which is based on quantum mechanics ought to be silenced by the memory of the reality of the events of Hiroshima and Nagasaki'.  Compared to solipsism, the proposition that relativity is not fundamental, and that 
the world is nonlocal, seems the lesser of two evils. This was certainly Bell's position, and is even seen as inevitable by some philisophers; Maudlin says \cite{Mau94}  
`I have argued that the [following is] unequivocal: Violation of
Bell's inequality can be accomplished only if there is superluminal
information transmission'.  

\section{Conclusion}

The centrality  of Bell's theorem to the foundations of modern physics is undeniable. 
It forces us to choose between a nonlocal interpretation of QM (either well-defined like the Bohmian interpretation, or ill-defined like Bohr's Copenhagen interpretation), and extreme subjectivism. 
But what about Einstein's theorem, as I have called it? Is it, as Einstein's biographer Pais thought \cite{Pai82}, only of interest for what it reveals about Einstein's state of mind? I suggest the contrary. 

For the majority of physicists who reject the idea of hidden variables, believing statistical QM to be complete,  Bell's theorem is logically superfluous. The much simpler theorem due to Einstein is logically sufficient to make such physicists face the choice between accepting that relativity is not fundamental, 
and denying that their colleagues, spouses {\em et cetera} have an independent existence. Perhaps if Einstein's theorem were better known, some of those physicists would be less hasty in their dismissal of hidden variable interpretations such as Bohm's. 

Cushing \cite{Cus94} has speculated about what would have happened if Bell's theorem had been 
discovered in the late 1920s, rather than 1964. This is a quite conceivable scenario, since the inequalities Bell used in his proof had in fact been formulated 
(as `conditions of possible experience') more than a century before Bell by the logician George Boole \cite{Boo1862}. Cushing argued that if Einstein had known that nonlocality of the uncontrollable kind was an inevitable consequence of QM plus reality, then he would have chosen 
reality over locality (even at the expense of his own postulates for the theory of relativity). 
In that case, he would have had no objections to de Broglie's 
theory, and the full development of the theory by Bohm from 1952 on might have happened almost 
a quarter of a century earlier. At this point, with the support of Einstein, in the absence of von Neumann's flawed theorem, and with the open-mindedness of Born and Jordan \cite{Dic98},  
what is now known as Bohmian mechanics might have been a serious contender with the Copenhagen interpretation to become the quantum orthodoxy. 

 \ack
This work was supported by the Australian Research Council.
I gratefully acknowledge the assistance and advice of Sheldon Chow, Damian Pope, Nadine Wiseman, Dave Kielpinski, Rob Spekkens, and Travis Norsen.

\begin{appendix}
\section*{Appendix: EPR versus Einstein} \label{Sec:app}
As noted in Sec.~\ref{sec:EPR}, Fine \cite{Fin96} and Norsen \cite{Nor05} have argued that Einstein's argument against the Copenhagen interpretation in 1935 was distinct from, and even superior to, that in the EPR paper. The EPR paper was actually written by Podolsky, and Einstein expressed his unhappiness  with the way it had turned out in a letter to \sch\  \cite{Fin96}. In this letter, Einstein presented his own argument, which is what Fine and Norsen contrast with the EPR argument. 

Einstein considers the EPR state, but then presents an argument in the following spirit \cite{Fin96}. For this state, a distant observer can find out the position of $S_2$ by a measurement on $S_1$. By locality, such a measurement cannot in any way disturb $S_2$. Thus, by EPR's definition, it follows that the position of $S_2$ is an element of physical reality. But Copenhagen QM says that the position  of $S_2$ is indeterminate, with an arbitrarily large uncertainty. Therefore Copenhagen QM is incomplete.  (The same argument could also be made using the momentum instead of the position.)

In the EPR paper,  Einstein's point could be deduced: that the position of $S_2$ is an element of reality, whereas  Copenhagen QM says it is indeterminate, and (incidentally) likewise for the momentum.  However Einstein thought that his point was obscured by an emphasis on the fact that {\em both} the position and the momentum were elements of reality according to EPR's definition, whereas in Copenhagen QM simultaneous values for these quantities are forbidden by  \hei's  uncertainty principle. This was a fact about which Einstein said `I couldn't care less' \cite{Fin96}. Maudlin \cite{Mau94} called it `an unnecessary bit of grandstanding' which muddied the argument by appearing to introduce counterfactual reasoning. 
 
While Einstein's argument is certainly worth contrasting with that in the EPR paper, 
it seems to me that Podolsky had in fact identified a crucial point. That is,  considering both 
position and momentum measurements is necessary if one wishes to turn Einstein's  thought-experiment into a proposal for a real experiment.

Recall the discussion in Sec.~\ref{sec:EPR} as to why the EPR state is superior to the single 
particle state in being able to show experimentally the nonlocality of the Copenhagen interpretation. 
The crucial point is that it is not possible to verify by local measurements on a single particle that its
state really is a pure superposition of transmitted and reflected, rather than a proper mixture.   By contrast, in the EPR scenario one could verify by local measurements that the two-particle state was pure, by verifying that the position correlations and momentum correlations were as predicted by QM.  
 
A pure state, and in particular the perfectly correlated pure state that EPR used, is an idealization. In any real experiment the correlations would not match those of the EPR state. What, then, would be sufficient experimental verification of the predictions of QM to enable Einstein's 1935 conclusion (that 
either orthodox QM is incomplete, or the principles of relativity are not valid) to be deduced? 
Consider just the status of the position $q_2$ of $S_2$. Say an observer can, by measuring $q_1$, determine $q_2$ with an accuracy $\sigma_q$. Then (assuming locality) this will prove that orthodox QM is incomplete as long as the state of $S_2$ is not a proper mixture of states with a position uncertainty $\delta_q \leq \sigma_q$. The reason is the same as for the single particle: if it were such a mixture then even in Copenhagen QM one could maintain that the position, to within an accuracy of $\sigma_q$, was an element of reality. Now if the state of $S_2$ were a proper mixture of states with $\delta_q \leq \sigma_q$ then, from \hei's uncertainty relation \cite{Hei27}, this would mean that it was a proper mixture of states that have momentum uncertainty $\delta_p \geq \hbar/2\sigma_q$. Thus, if an observer could, by measuring $p_1$, determine $p_2$ with an accuracy $\sigma_p$ less than this lower bound on $\delta_p$, that would prove that the state of $S_2$ was not such a proper mixture, as desired. Thus, the QM prediction that underlies Einstein's argument is that it is possible, by {\em different} measurements on $S_1$, to determine the position or the momentum of $S_2$ to within accuracies that satisfy $\sigma_q\sigma_p \leq \hbar/2$. That is, their product is smaller than the product of uncertainties that appears in the \hei\ uncertainty relation. This is precisely the condition that has been derived as being necessary to demonstrate the EPR paradox \cite{Rei89} in the form that Podolsky presented it. 

Thus, in the end, although Einstein's argument differs from the published EPR argument, they rely upon exactly the same physical correlations. Given that in later life Einstein seemed quite happy to present the argument essentially as it appeared in the EPR paper \cite{Ein46}, I suggest that the distinction between Einstein's 1935 argument and the EPR argument is not one that deserves to be overly emphasized.

\end{appendix}

\section*{Author Information}

Howard Wiseman did his B.Sc.~(1991) and Ph.D.~(1994) at the University of Queensland, 
followed by a post-doc at the University of Auckland. Since then he has held 
fellowships from the Australian Research Council, being currently Federation Fellow 
and Professor at Griffith University. His principle research areas are quantum feedback control, quantum information, and fundamental questions in quantum mechanics.
The current article arose from a need to understand the background and implications
of Bell's theorem, for satisfying personal curiosity, for motivating other research, and for teaching  
non-physicists.

\section*{References}
\begin{thebibliography}{10}
\bibitem{Lan74}
C. Lanczos, 
{\em The Einstein Decade (1905--1915)}
(Academic Press, New York, 1974).

\bibitem{EinPodRos35}
A. Einstein, B. Podolsky, and N. Rosen, `Can quantum-mechanical
description of physical reality be considered complete?', Phys.
Rev. {\bf 47}, 777--780 (1935).

\bibitem{Ein46}
A. Einstein, `Autobiographical Notes', in Ref.~\cite{Sch49}, 
pp.~1--94, with translation (the odd-numberred pages) by P. A. Schilpp. 

\bibitem{Sch49}
P. A. Schilpp (ed.), {\em Albert Einstein: Philosopher-Scientist}
(Library of the Living Philosophers, Evanston, 1949).

\bibitem{PleVed98}
M. B. Plenio and V. Vedral,
`Teleportation, entanglement and thermodynamics in the 
quantum world', Contem. Phys. {\bf 39}, 431--446 (1998).

\bibitem{NieChu00}
M. A. Nielsen and I. L. Chuang, 
{\em Quantum Computation and Quantum Information}
(Cambridge University Press, 2000).

\bibitem{SpiMunBarKok05}
T. P. Spiller, W. J. Munro, S. D. Barrett, and P. Kok,
`An introduction to quantum information processing:
applications and realizations',
Contemp. Phys., {\bf 46}  407--436 (2005).

\bibitem{Red05}
S. Redner, 
`Citation Statistics from 100 Years of Physical Review',
Physics Today {\bf 58}:6, pp.~49--54 (2005).

\bibitem{Sch35}
E. Schr\"odinger, `The present situation in quantum mechanics',
Naturwissenschaften {\bf 23}, 823--881 (1935); English translation
by J. D. Trimmer in Proc. Am. Phil. Soc. {\bf 124}, 323--338 (1980).

\bibitem{Bel64}
J. S. Bell, `On the Einstein Podolsky Rosen Paradox', Physics {\bf
1}, 195--200 (1964). Reproduced in Ref.~\cite{BellCollection}.

\bibitem{BellCollection}
M. Bell, K. Gottfried, and M. Veltman (eds.), {\em John S. Bell on
the Foundations of Quantum Mechanics} (World Scientific, Singapore,
2001).

\bibitem{Har98}
L. Hardy, `Spooky action at a distance in quantum mechanics', 
Contemp. Phys. {\bf 39}, 419--429 (1998).

\bibitem{JacWis05}
K. A. Jacobs and H. M. Wiseman, `An entangled web of crime: Bell's
theorem as a short story', to be published in Am. J. Phys. ({\em Preprint} quant-ph/0504192).

\bibitem{Bel99}
M. Beller, {\em Quantum Dialogue: The Making of a Quantum
Revolution} (University of Chicago Press, Chicago, 1999).

\bibitem{Hei27}
W. Heisenberg, `On the physical content of quantum theoretical kinematics and mechanics',
Z. f\"ur
Physik {\bf 43}, 172--198 (1927). English translation in
Ref.~\cite{WheZur83}.

\bibitem{WheZur83}
J.A. Wheeler and W.H. Zurek (eds.)
{\em Quantum Theory and Measurement}
(Princeton, New Jersey, 1983).

\bibitem{Boh28}
N. Bohr, `The quantum postulate and the recent development of
atomic theory', Nature {\bf 121}, 580--590 (1928). Reproduced in
Ref.~\cite{WheZur83}.

\bibitem{Wic95}
D. Wick, {\em The Infamous Boundary: Seven Decades of Controversy in
Quantum Physics} (Birkhauser, Boston, 1995).


\bibitem{deB56}
L. de Broglie, 
{\em Non-linear Wave Mechanics: A Causal Interpretation}  (Elsevier,
Amsterdam, 1960).

\bibitem{Jam74}
M. Jammer, {\em The Philosophy of Quantum Mechanics}, 
(John Wiley and Sons, New York, 1974).

\bibitem{Cus94}
J. T. Cushing, {\em Quantum Mechanics: Historical Contingency and
the Copenhagen Hegemony} (The University of Chicago Press, Chicago,
1994).

\bibitem{Bel97}
D. W. Belousek, `Einstein's 1927 unpublished hidden-variable theory:
It's background, context and significance', \emph{Stud. Hist. Phil.
Mod. Phys.} {\bf 27}, 437--461 (1997).

\bibitem{Val01}
A. Valentini,
{\em Pilot-Wave Theory: An Alternative Approach to Modern Physics}
(Cambridge University Press, Cambridge, forthcoming).

\bibitem{Hei30}
W. Heisenberg,
{\em The Physical Principles of Quantum Mechanics}
(The University of Chicago Press, Chicago, 1930).

\bibitem{Ein05}
A. Einstein 
`On the Electrodynamics of Moving Bodies', 
in  {\em The Principle of Relativity} (Methuen, London, 1923) --- 
a translation of the paper in Annalen der Physik. {\bf 17}, 891--921 (1905).

\bibitem{BohrEinst} N. Bohr, `Discussion with Einstein on epistemological problems in atomic phyiscs', in Ref.~\cite{Sch49}, pp.~201--241; reproduced in \cite{WheZur83}.

\bibitem{Wig32}
E. P. Wigner, `On the quantum correction for thermodynamic
equilibrium', Phys. Rev. {\bf 40}, 749--759 (1932).

\bibitem{Pai82}
A. Pais,
{\em Subtle is the Lord}
(Oxford U. Press, Oxford, 1982).

\bibitem{Ein49}
A. Einstein, `Rely to Critics', 
in Ref.~\cite{Sch49}, pp.~663--688 (1949).

\bibitem{Fin96}
A. Fine, {\em The
Shaky Game}, (University of Chicago Press, Chicago, 1996).

\bibitem{Nor05}
T. Norsen, `Einstein's boxes', Am. J. Phys. {\bf 73}, 164--76 (2005).

\bibitem{dEs89}
B. d'Espagnat, {\em Conceptual Foundations of Quantum Mechanics} 2nd Ed.
(Addison-Wesley, New York, 1989).

\bibitem{Rei89}
M. D. Reid, 
`Demonstration of the Einstein-Podolsky-Rosen paradox using nondegenerate parametric amplification',
Phys. Rev. A {\bf 40}, 913--923 (1989).

\bibitem{Boh35}
N. Bohr, `Can quantum-mechanical description of physical reality be
considered complete?', Phys. Rev. {\bf 48}, 696--702 (1935).

\bibitem{Ros64}
L. Rosenfeld, `Neils Bohr in the thirties', in S. Rozental (ed.),
{\em Neils Bohr: His life and times as seen by his friends and
colleagues} (North Holland, Amsterdam, 1964), pp. 114--136; English
edition 1967.

\bibitem{Boh63} 
N. Bohr, {\em Essays 1958--62 on Atomic Physics and Human Knowledge}
(Wiley, New York, 1963).

\bibitem{dEs84}
B. d'Espagnat, `Nonseparability and the tentative descriptions of
reality', Physics Reports {\bf 110}, 202--263 (1984).

\bibitem{Cus94b}
J. T. Cushing, `Locality / separability: Is this necessarily a
useful distinction?', in {\em PSA: Proceedings of the Biennial
Meeting of the Philosophy of Science Association}, Volume One:
Contributed Papers, pp. 107--116 (1994).

 \bibitem{Sch35b} 
 E. Schr\"odinger, `Discussion of Probability Relations Between
Separated Systems,Ó Proc. Camb. Phil. Soc. {\bf 31},
553--563 (1935). %

\bibitem{Sch36}
E. Schr\"odinger,
 `Probability relations between separated systems',
 Proc. Camb. Phil. Soc. {\bf 32}, 446--452 (1936).
 
\bibitem{Von32}
J. von Neumann, {\em Mathematical Foundations of Quantum Mechanics}
(Springer, Berlin, 1932);
English translation (Princeton University Press, Princeton, 1955).

\bibitem{Fey95}
P. K. Feyerbrand, {\em Killing Time} (Chicago U. Press, Chicago, 1995). 
 
\bibitem{Boh52}
D. Bohm, `A suggested interpretation of the quantum theory in terms
of `hidden' variables' (parts I and II) Phys. Rev. {\bf 85}, 166--193
(1952).

\bibitem{BohHil93}
D. Bohm  and B. J. Hiley, {\em The Undivided Universe: An
Ontological Interpretation of Quantum Theory} (Routledge, London,
1993).

\bibitem{Whi96}
A. Whitaker, {\em Einstein, Bohr and the Quantum Dilemma} (Cambridge
University Press, Cambridge, 1996).

\bibitem{Boh51}
D. Bohm, {\em Quantum Theory} 
(Prentice-Hall, New York, 1951). 

\bibitem{BohAha57}
D. Bohm and Y. Aharonov,
`Discussion of Experimental Proof for the Paradox of Einstein, Rosen, and Podolsky',
Phys. Rev. {\bf 108}, 1070--1076 (1957).

\bibitem{Nor04}
T. Norsen, 
`EPR and Bell Locality', ({\em Preprint}  quant-ph/0408105).

\bibitem{How85}
D. Howard, `Einstein on Locality and Separability'.
Stud. Hist. Phil. Sci. {\bf 16}, 171-201 (1985).

\bibitem{Hei58}
W. Heisenberg,
{\em Physics and Philosophy}
(Allen \& Unwin, London, 1958).
 
 \bibitem{Hei58b}
W. Heisenberg, `The Representation of Nature in Contemporary Physics', 
Daeladus {\bf 87}, 95--108 (1958), p.~100. 

\bibitem{Eve57}
H. Everett III, ` `Relative state' formulation of quantum
mechanics' Rev. Mod. Phys. {\bf 29}, 454--462 (1957).
 
\bibitem{Bel66}
J. S. Bell, `On the problem of hidden variables in quantum
mechanics', Rev. Mod. Phys. {\bf 38}, 447--452 (1966). Reproduced
in Ref.~\cite{BellCollection}. 

\bibitem{Sta71} H. P. Stapp, `S-matrix interpretation of quantum theory', Phys. Rev. D {\bf 3}, 1303--1320 (1971).

\bibitem{Ebe77}
P. H. Eberhard, `Bell's theorem without hidden variables', Il
Nuovo Cimento {\bf 38B}, 75--80 (1977).

\bibitem{Bel81}
J. S. Bell, `Bertlmann's socks and the nature of reality', Journal
de Physique, Colloque C2, suppl. au numero 3, Tome 42, 41--61
(1981). Reproduced in Ref.~\cite{BellCollection}.

\bibitem{Fra82}
B. C. van Fraasen, `The charybdis of realism: Epistemological
implications of Bell's inequality', Synthese {\bf 52}, 25--38
(1982). 

\bibitem{Pri96}
H. Price, {\em Time's Arrow and Archimedes' Point} (Oxford
University Press, Oxford, 1996).

\bibitem{Bub74}
J. Bub, {\em The Interpretation of Quantum Mechanics} (Reidel,
Dordrecht, 1974).

\bibitem{Bel77}
J. S. Bell, `Free variables and local causality', Epistemological
Letters, Feb. (1977). Reproduced in Ref.~\cite{BellCollection}.

\bibitem{Bel90}
J. S. Bell, `Against Measurement', 
Physics World {\bf 3}, 33 (1990) (August). Reproduced in Ref.~\cite{BellCollection}.

\bibitem{Bel87}
J. S. Bell, {\em Speakable and Unspeakable in Quantum Mechanics},
(Cambridge University Press, Cambridge, 1987).

\bibitem{Bel81a}
J. S. Bell, `Quantum Mechanics for Cosmologists', 
in {\em Quantum Gravity 2}, C. Isham, R. Penrose, and D. Sciama 
(Oxford University, Oxford, 1981), pp.~611--637.  Reproduced in Ref.~\cite{BellCollection}.

\bibitem{Wig76}
E. P. Wigner, `Interpretation of Quantum Mechanics',
(1976), reprinted in Ref.~\cite{WheZur83}.

\bibitem{Mer93}
N. D. Mermin, 
'Hidden Variables and the Two Theorems of John Bell', 
Rev. Mod. Phys. {\bf 65}, 803--815 (1993). 

\bibitem{Bel80}
J. S. Bell, 
`de Broglie-Bohm, Delayed Choice, Double-Slit Experiment, and Density Matrix', 
Int. J. Quantum Chem.: Quantum Chem. Symposium {\bf 14}, 155-159 (1980). 
Reproduced in Ref.~\cite{BellCollection}.

\bibitem{AspGraRog81}
       A. Aspect, P. Grangier, and G. Roger,
       `Experimental tests of realistic local theories via Bell's theorem',
       Phys. Rev. Lett. {\bf 47}, 460--463 (1981); {\em ibid.}
       `Experimental realization of Einstein-Podolsky-Rosen-Bohm {\em Gedankenexperiment}:
       A new violation of Bell's inequalities', Phys. Rev. Lett.
       {\bf 49}, 91--94 (1982).

\bibitem{Red87}
M. Redhead, {\em Incompleteness, Nonlocality, and Realism} (Clarendon, Oxford, 1987), pp. 107--113,

\bibitem{Shi84}
A. Shimony, `Controllable and uncontrollable non-locality', in S.
Kamefuchi {\em et al.} (eds.), {\em Foundations of Quantum Mechanics
in the Light of New Technology} (Physical Society of Japan, Tokyo,
1984), pp. 225--230.

\bibitem{Shi93}
A. Shimony, {\em Search for a Naturalistic World View} 
(Cambridge, UK, 1993).  

\bibitem{Jar84}
J. Jarrett, (1984), `On the physical significance of the locality conditions in the Bell argument', No\^us {\bf 18}, 569--89.

\bibitem{Bra04}
   G. Brassard, A. Broadbent, and A. Tapp, `Quantum pseudo-telepathy', Found. Phys. {\bf 35}, 1877--1907 (2005). 

\bibitem{Mau94}
T. Maudlin, {\em Quantum Non-Locality and Relativity} (Blackwell,
Oxford, 1994).

\bibitem{Ken05}
A. Kent,
`Causal Quantum Theory and the Collapse Locality Loophole',
Phys. Rev. A {\bf 72}, 012107 (2005).

\bibitem{Sta77} 
H. P. Stapp, `Are superluminal connections necessary?', Il Nuovo
Cimento {\bf 40B}, 191--205 (1977).


\bibitem{OED}
The New Shorter Oxford English Dictionary (1993).

\bibitem{Pop85}
K. Popper,
{\em Quantum Theory and the Schism in Physics}
(Routledge, London, 1985). 

\bibitem{Boo1862}
G. Boole, `On the theory of probabilities', Philosophical
Transactions of the Royal Society of London {\bf 152}, 225--252
(1862). See also G. Boole, {\em The Laws of Thought} (Dover, New
York, 1958).

\bibitem{Dic98}
W. M. Dickson, {\em Quantum Chance and Nonlocality}
(Cambridge, UK, 1998).


\end{thebibliography}
\end{document}